# Gaia astrometric and photometric study of open clusters Dolidze 18 & Ruprecht 70

Tadross, A. L.[a] (altadross@gmail.com) and Hendy, Y. H.[a]

[a]National Research Institute of Astronomy and Geophysics, 11421 - Helwan, Cairo, Egypt

**Abstract**

Here, we conducted a photometric and astrometric study of two open stellar clusters Dolidze 18 and Ruprecht 70, which have not been photometrically studied before. The most important thing for using Gaia (DR2) database lies in the positions, parallax, and proper motions, which make us split cluster members from the field ones and get precise astrophysical parameters. On studying the radial density profiles of these clusters, the actual sizes are estimated and found larger using Gaia. From the color-magnitude diagrams and theoretical isochrones, we simultaneously determined the ages, distance moduli, and reddening of the two clusters. However, considering the parallaxes of Gaia (DR2) for the cluster members, we calculated the cluster distance and confirmed what we obtained from the color-magnitude diagram. Then, the Cartesian galactocentric coordinates ($X_\odot$, $Y_\odot$, $Z_\odot$), and the distances from the galactic center ($R_g$) were also estimated. According to luminosity and mass functions, the total luminosity and total mass of the clusters are estimated. Our study shows that Ruprecht 70 is recently dynamically relaxed, while Dolidze 18 is not relaxed yet.

## 1. Introduction

Open star clusters (OCs) are stellar systems that originated under the same astrophysical environments, at the same distance, with identical ages, original chemical compositions, and various masses. These objects are crucial to explaining the Milky Way Galaxy's structure and evolution (cf. Gilmore et al. (2012); Moraux. (2016)). Our data on OCs came from well-known global catalogs and databases, such as Dias et al. (2002, 2014) and Webda (https://webda.physics.muni.cz/navigation.html), which provide all the knowledge required to conduct accurate studies of such objects.

In global catalogs, there are around 3000 known open star clusters. Maybe there are a lot of unknown OCs in our Milky Way Galaxy (cf. Piskunov et al. (2006); Portegies Zwart et al. (2010); Martinez-Medina et al. (2016); Cantat-Gaudin et al. (2020)). We believe that most of them would be detected in the Gaia era. Gaia (DR2) is a high-precision database containing 1.7 billion references in the astrometric parameters of coordinates, proper motions, and parallaxes ($\alpha$, $\delta$, $\mu\alpha \cos\delta$, $\mu\delta$, $\pi$). Furthermore, magnitudes for over 1.3 billion references in three photometric filters (G, $G_{BP}$, $G_{RP}$), Gaia Collaboration et al. (2018). The Gaia Archive can be accessed via the website (http://ww.cosmos.esa.int/gaia). As a result, we can distinguish cluster members from field star contaminants with greater precision. Using the Gaia (DR2) database, Bossini et al. (2019) studied the age, distance moduli, and extinction of 269 open clusters. Cantat-Gaudin et al. (2020) used the Gaia (DR2) database to calculate the membership probabilities of 2000 OCs, but the clusters under study were not included. We have conducted the first Gaia-based photometric study of these two clusters.

The objects Dolidze 18 and Ruprecht 70 have few astrometric studies, e.g., Dolidze (1961); Ruprecht (1966); Koposov et al. (2008); Dias et al. (2014). Our goal is to estimate the main parameters of the two poorly studied clusters using the most recent database of the Gaia mission.

This paper is organized as follows: the data analysis is presented in Section 2. The radial density profiles (RDP), the clusters' centers, and diameters are depicted in Section 3. The color-magnitude diagrams (CMD) and photometry are performed in Section 4. Luminosity functions (LF), mass functions (MF), and the dynamical status of the clusters (the relaxation time $T_R$) are described in Section 5. Finally, the conclusions of our study are summarized in Section 6.

## 2. Data Analysis

Using the world website SIMBAD (http://simbad.u-strasbg.fr/simbad/) we can get the central coordinates of the two clusters Dolidze 18 and Ruprecht 70. These clusters lie in constellations Auriga and Carina, respectively. The source data downloaded from the Gaia (DR2) database service at the given centers with radii of 25 and 15 arcmin, respectively. Fig. 1 represents the optical images of the clusters under study as taken from the world-known tool ALADIN (https://aladin.ustrasbg.fr/AladinLite) at DSS colored optical wavelength and field of view (FOV) equals 12 arcmin for both.

According to Lindegren et al. (2018), all the parallax values should be shifted by adding 0.03 mas to their values. In addition, RUWE, the value for a renormalized unit weight error, which indicates how well the source matches the single-star model; RUWE should be < 1.4. The parallax uncertainty ranges from 0.04 mas for sources at G < 15 mag, to approximately 0.1 mas at G = 17 mag and up to 0.7 mas at G = 20 mag. The corresponding uncertainty in the respective proper motion components acts from 0.06 mas/yr (for G < 15 mag) to about 0.20 mas/yr (for G = 17 mag) and up to 1.2 mas/yr (for G = 20 mag).

Both parallax and proper motion are also affected by systematic errors, on the order of 0.1 mas and 0.1 mas/yr, respectively.

Actually, we used sources brighter than 20-mag in the G band. In addition, the parallaxes errors and proper motions errors for both clusters are taken at < 0.7 mas and < 1.0 mas/yr, respectively. Undoubtedly, issues such as binarity and missing stars can also affect the results, especially the luminosity and mass functions. Extending the magnitude limit and applying such a threshold is to enhance, of course, the inferred results.

The cluster members have shared similar astrometrical properties. Thus, the vector point diagrams (VPD) of proper motion components (pmRA and pmDE) are plotted for each cluster, as shown in Fig. 2. The highest density area is chosen as a subset of the most probable members of the cluster. A circle has drawn around the most darkened region we selected, from which we can define the co-moving stars, Tadross (2018) and Tadross & Hendy (2021). Co-moving stars are stars that move together with the same speed and the same direction in the sky compared to the background field ones, see Fig. 3. This is one of the conditions of membership that we have applied, besides the parallax, proper motions, the cluster boundary, and the CMD fitting, all of which are taken into account.

The cluster parallax range can be seen as shown in the Lefthand panels of Fig. 4. The relation between the magnitudes and parallax shows the concentration of the cluster members, while the stars of the field seem to be dispersed, as shown in the right-hand panels of Fig. 4. In this respect, a star is counted as a cluster member when located within the clusters size (see section 3), lies within the parallax and proper motion ranges of the cluster, and having the same velocity in the sky compared to the background field stars.

## 3. Clusters structure (centers and diameters)

The cluster center is taken at the most over-density area of the cluster region. Hence, using the database of Gaia, we divided the area of the cluster into equal-sized bins in RA and DE and counting the stars in each bin. The gaussian fitting profile has been applied to both directions, as shown in Fig. 5. The obtained central coordinates of Dolidze 18 and Ruprecht 70 are found to be in good agreement with Dias et al. (2002) Catalog, see Table 1.

To obtain the diameters of the studied clusters, we divided the clusters area into concentric circles (shells), each having a radius of 1 arcmin bigger than the previous one. The density of stars in each shell is calculated. The density distributions of the candidate clusters show good fits with the empirical King (1966) model that can be presented as:

$$f(R) = f_{bg} + \frac{f_0}{1+(R/R_c)^2}$$

where ($f_{bg}$) is the background stellar density, ($f_o$) is central density, and ($R_c$) is the core radius of the cluster, which is the distance at which the stellar density equals half the central density. The limited radius ($R_{lim}$) can be estimated at that value, which covers the whole cluster area and reaching sufficient constancy, where the cluster density melted in the background field stars, see Fig. 6. $f_{bg}$, $f_o$, $R_c$, and $R_{lim}$ are given in Table 1. On the other hand, the tidal radius of a cluster, which is the distance from the cluster core, at which the gravitational impact of the Galaxy equivalents to that of the cluster core. Calculating the total masses of Dolidze 18 and Ruprecht 70 (as mentioned in Sec. 5), the tidal radius can be calculated for both clusters using Jeffries et al. (2001) equation:

$$R_t = 1.46 \, x \, (Mc)^{1/3}$$

where R t is the tidal radius (in parsec), and Mc is the total mass of the cluster (in solar mass). According to Peterson and Peterson & King (1975), the concentration parameter $C = \log (R_{lim}/R_c)$ shows us how the cluster is prominent in comparison with the background stars. It is found to be C = 1.07 and 1.37. Therefore, we found that Dolidze 18 does not appear to have star condensation enough at its center.

## 4. Color-Magnitude Diagram

CMD is an essential tool for estimating the fundamental parameters of the candidate clusters. Age, distance modulus, color excess, and metallicity can be estimated all at once by applying one of the Padova stellar isochrones and obtaining the best fit to the observed CMD. It is substantial to recognize the sequence of the cluster stars from the field ones because the stars in the cluster region are contaminated with background field. We used the Padova PARSEC (http://stev.oapd.inaf.it/cgi-bin/cmd) version 1.2S database of stellar evolutionary tracks and isochrones of Bressan et al. (2012), which uses the Gaia filter passbands of Evans et al. (2018). It scaled to the solar metallicity Z = 0.0152. The best fit was obtained from the visual isochrone fit, where three suitable isochrones of different ages have been applied to each cluster as shown in Fig. 7. The main parameters of Dolidze 18 and Ruprecht 70 are verified with ages of 20±10 and 50±10 Myr. The visual reddening and the intrinsic distance moduli are found to be 0.62±0.11 & 12.95±0.60, and 0.45±0.08 & 11.65±0.45 mag, respectively. To ensure accuracy, we applied the ASteCA (Automated Stellar Cluster Analysis) code (https://asteca.readthedocs.io/en/latest/), following all installation instructions (cf. Perren et al. (2015, 2020)). We de-redden the colors and magnitudes of the two clusters and used them as input values for the ASteCA code. The main results of the clusters are consistent well with our visual fit, where the ages are found to be 16.5±10 and 56±10 Myr for Dolidze 18 and Ruprecht 70, respectively. The visual reddening and intrinsic distance moduli are found to be 0.58±0.07 & 12.70±0.47, and 0.47±0.09 & 11.66±0.35 mag, respectively. It is worthy to note that Dolidze 18 is close to a heavy dusted area in the Galaxy and hence the field has a strong differential extinction, as shown in the 3D extinction map along this line of sight (http://argonaut.skymaps.info).

We used the absorption ratio of the photometric systems in different wavelengths to the visual absorption ($A_\lambda/A_v$), Cardelli et al. (1989); O'Donnell. (1994); and the Padova website CMD 3.5 (http://stev.oapd.inaf.it/cgi-bin/cmd). The absorption ratios of Gaia (DR2) are $A_G/A_v = 0.861$, $A_{BP}/A_v = 1.062$, and $A_{RP}/A_v = 0.651$. These ratios have been used for correction of the magnitudes for interstellar reddening and converting the color excess to $E(B - V)$, where $R_v = A_v/E(B - V) = 3.1$ and $E(B - V) = 0.785 \, E(BP - RP)$.

The parallax values of Gaia (DR2) are affected by an offset of -0.03 mas, Lindegren et al. (2018), which has been considered here. However, depending on the mean parallax of

Dolidze 18 and Ruprecht 70, yield distances of 3775 and 2220 pc, respectively. They agree well with the CMD fitting distances of the two clusters. The members of the cluster were selected for the stars, which are inside the diameter of the cluster and lie in the parallax and proper motion valid ranges. Correspondingly, the Cartesian galactocentric coordinates (X, Y, Z), and the distances from the galactic center (Rg) are estimated for the two clusters and listed in Table 1 (cf. Tadross (2011)). The Y-axis connects the Sun to the Galactic center, it is positive toward the Galactic anti-center, while the X-axis is perpendicular to Y-axis, it is positive in the first and second Galactic quadrants (Lynga (1982)).

## 5. Luminosity & mass functions, and dynamical state

A large number of stars of the same age and chemical structure but differing masses make up an open cluster. The luminosity and mass functions of a cluster are determined by the number of confirmed membership (LF & MF). The most difficult aspect of studying the LF and MF is removing field star pollution from the cluster's area. We can distinguish cluster members from field stars by using the cluster's true dimension, parallax, and proper motion conditions.

A high-degree polynomial equation between the absolute magnitudes and the masses of the main sequence stars can be derived from the isochrone that better fits the cluster. The distance modulus of the cluster is used to transform the apparent magnitudes of the cluster's members to absolute magnitudes.

Following that, the potential masses can be determined. In the left-hand panels of Fig. 8, a histogram of LF for each cluster has been plotted. We found that the brightest, most massive stars are clustered in the cluster's center, while the fainter and less massive stars are scattered outside. The total luminosity is found to be -8.7 and -5.8 mag for Dolidze 18 and Ruprecht 70, respectively. Based on the theoretical data of the applied isochrones, the members of the cluster are divided into mass intervals, which are linked to their absolute magnitude bins.

The resulting MF of Dolidze 18 and Ruprecht 70 are shown in the right-hand panels of Fig. 8. The linear fit represents the initial mass function (IMF) slope obtained from the following equation:

$$\frac{dN}{dM} \propto M^{-\alpha}$$

where dN/dM is the number of stars in the mass interval [M:(M+dM)] and α is the slope of the relation, which is found to be -1.20 and -2.40, respectively. It indicates that the MF slope of Ruprecht 70 agrees well with Salpeter (1955) value, while the MF slope of Dolidze 18 is small relative to the Salpeter value (it indicates that most members of the cluster < 0.5 $M_\odot$). The total masses of the target clusters are calculated by integrating the masses of the cluster members (Sharma et al. (2006)). They are found to be 800 and 325 $M_\odot$, respectively. Finally, the relaxation time ($T_R$) of the two studied clusters has been determined, where $T_R$ is the time scale when the cluster loses all traces of its initial conditions, i.e., the time for the cluster to reach the equivalent level of energy. It is given by the formula of Spitzer et al. (1971) as:

$$T_R = \frac{8.9 \times 10^5 \sqrt{N} \times R_h^{1.5}}{\sqrt{\langle m \rangle} \times log(0.4N)}$$

where N is the number of cluster members, $R_h$ is the radius containing half of the cluster mass (in parsecs; assuming that R h equals half of the cluster radius), and <m> is the average mass of the star in the cluster. It is a quotient of the total mass of the cluster and the number of members. Then, $T_R$ is found to be 55 and 14 Myr for Dolidze 18 and Ruprecht 70, respectively. According to the cluster age and its relaxation time, we can define the dynamical evolution parameter τ, where τ= Age / $T_R$. The age of the cluster should be sufficiently greater than the relaxation time. This parameter indicates that Ruprecht 70 is recently dynamically relaxed, while Dolidze 18 is not relaxed yet.

## 6. Conclusions

We present here a new Gaia (DR2) analysis of the two open clusters Dolidze 18 and Ruprecht 70 that have never been photometrically studied before. These clusters lie in constellations Auriga and Carina, respectively. We estimated the key physical parameters of the two clusters using the Gaia (DR2) database. Table 1 summarizes the physical parameters determined in this analysis. We used the ASteCA code to ensure precision and were able to achieve more precise parameters, although the actual sizes were found to be bigger. On studying the dynamical state, we found that Ruprecht 70 is recently relaxed, while Dolidze 18 is not relaxed yet. We suggested that Dolidze 18 may need a further study using 2MASS observations where it would be much less sensitive to extinction and could provide more valuable results.

## Acknowledgments

This work is a part of a project named IMHOTEP program No. 42088ZK between Egypt and France. We are grateful to the Academy of Scientific Research and to the Service of scientific cooperation in both countries for allowing us to work through this project and provide the necessary support to accomplish our scientific goals. This work has used data from the European Space Agency (ESA) mission Gaia processed by (www.cosmos.esa.int/web/gaia/dpac/consortium), i.e., the Gaia Data Processing & Analysis Consortium (DPAC). Funding for the DPAC has been provided by national institutions, in particular, the institutions participating in the Gaia Multilateral Agreement.

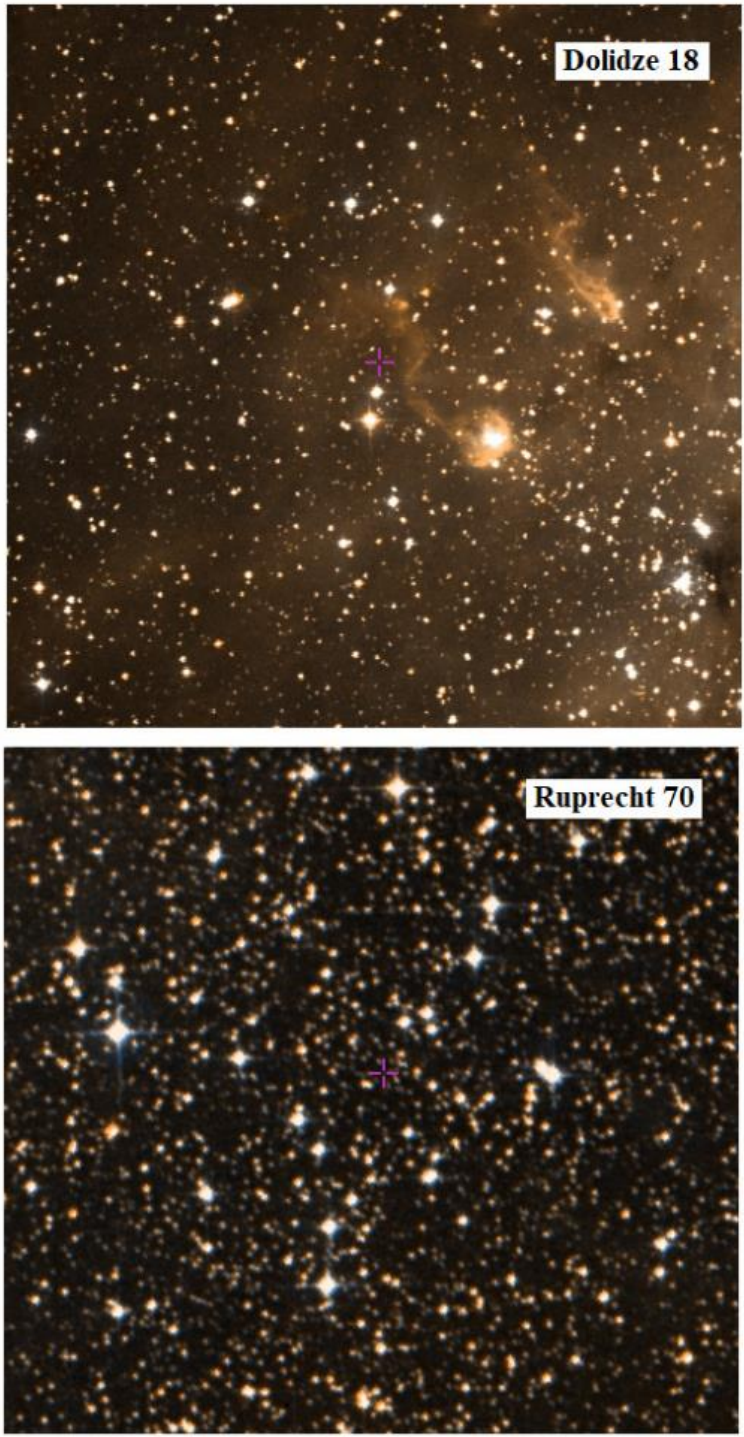

Fig. 1. The images of the clusters Dolidze 18 and Ruprecht 70 as taken from ALADIN at DSS colored optical wavelength and field of view (FOV) of 12 arcmin that corresponds to distances of 13.6 pc and 7.5 pc for Dolidze 18 and Ruprecht 70, respectively. These clusters lie in constellations Auriga, and Carina respectively. North is up, East is to the left.

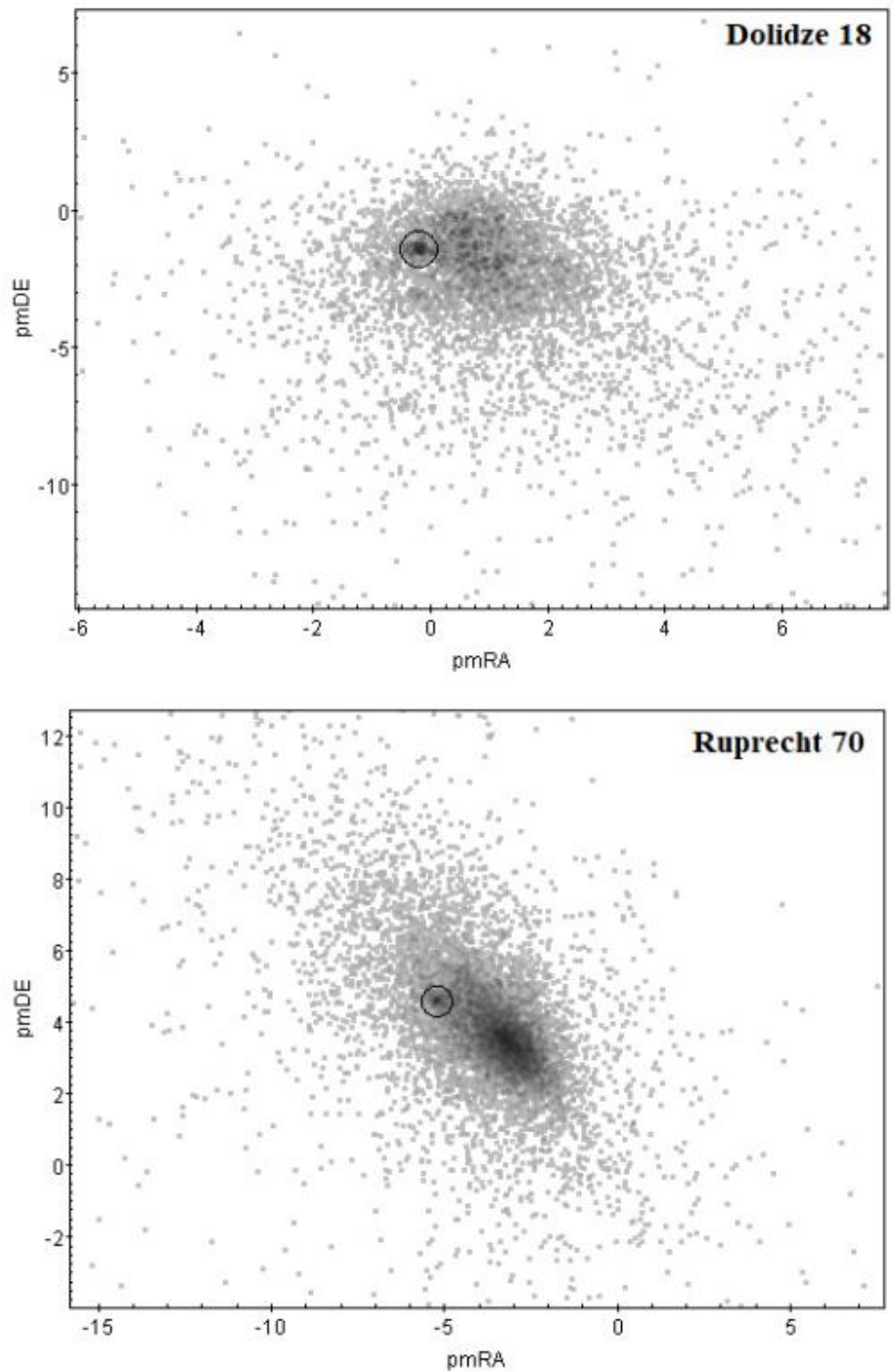

Fig. 2. Vector point diagrams of the clusters Dolidze 18 and Ruprecht 70, from which the highest concentrated area is selected as a subset of the most probable members of each cluster. The circles refer to the most darkened areas, where the stars are selected for the present study.

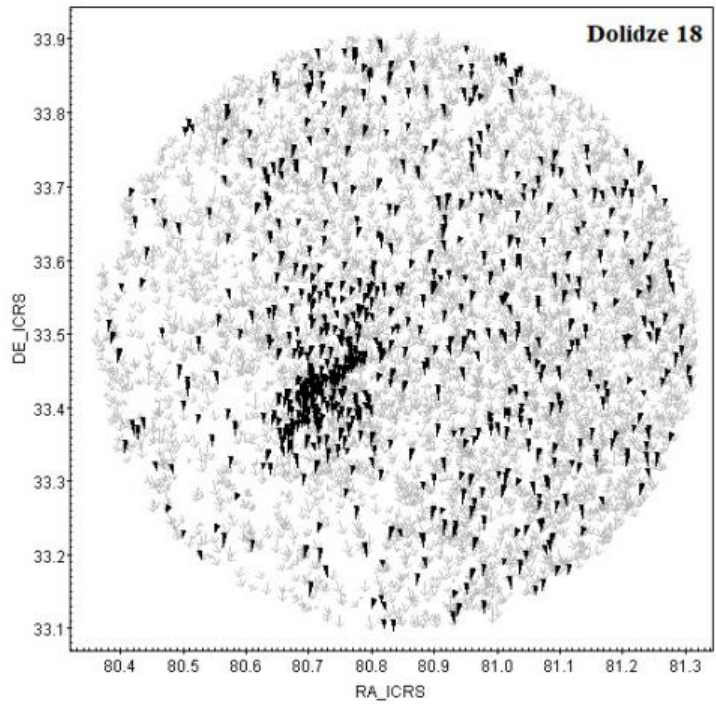

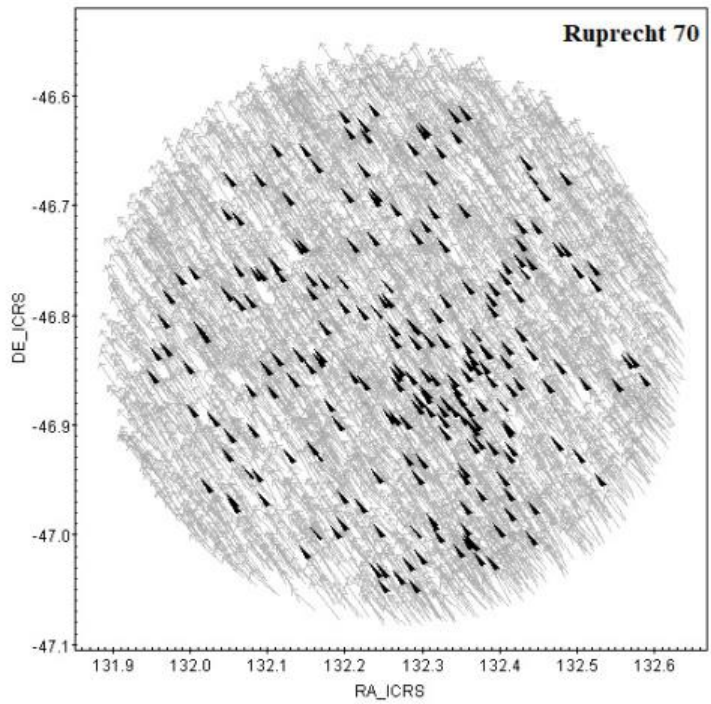

Fig. 3. The comoving stars of the clusters Dolidze 18 and Ruprecht 70. They represent the subsets that were selected from the VPDs. The black triangles refer to the clusters members, i.e., the stars that move together with the same speed and the same direction in the sky. While the gray arrows refer to the field stars that move at different speeds and in different directions in the sky.

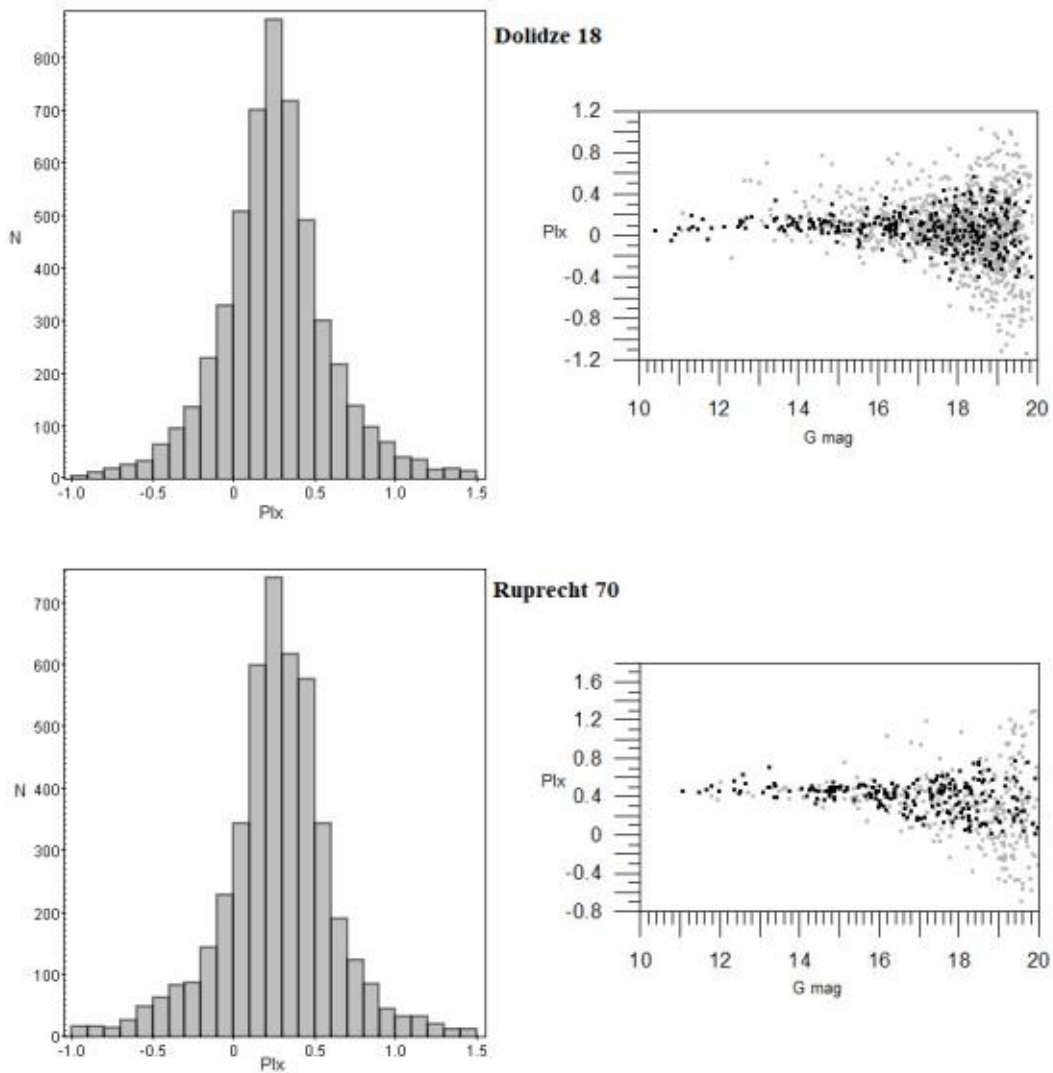

Fig. 4. The left-hand panels show the parallax ranges of the clusters Dolidze 18 and Ruprecht 70. The right-hand panels present the relation between the magnitude and parallax of each cluster. It shows concentration for the members of the cluster, while the stars in the field seem to be dispersed.

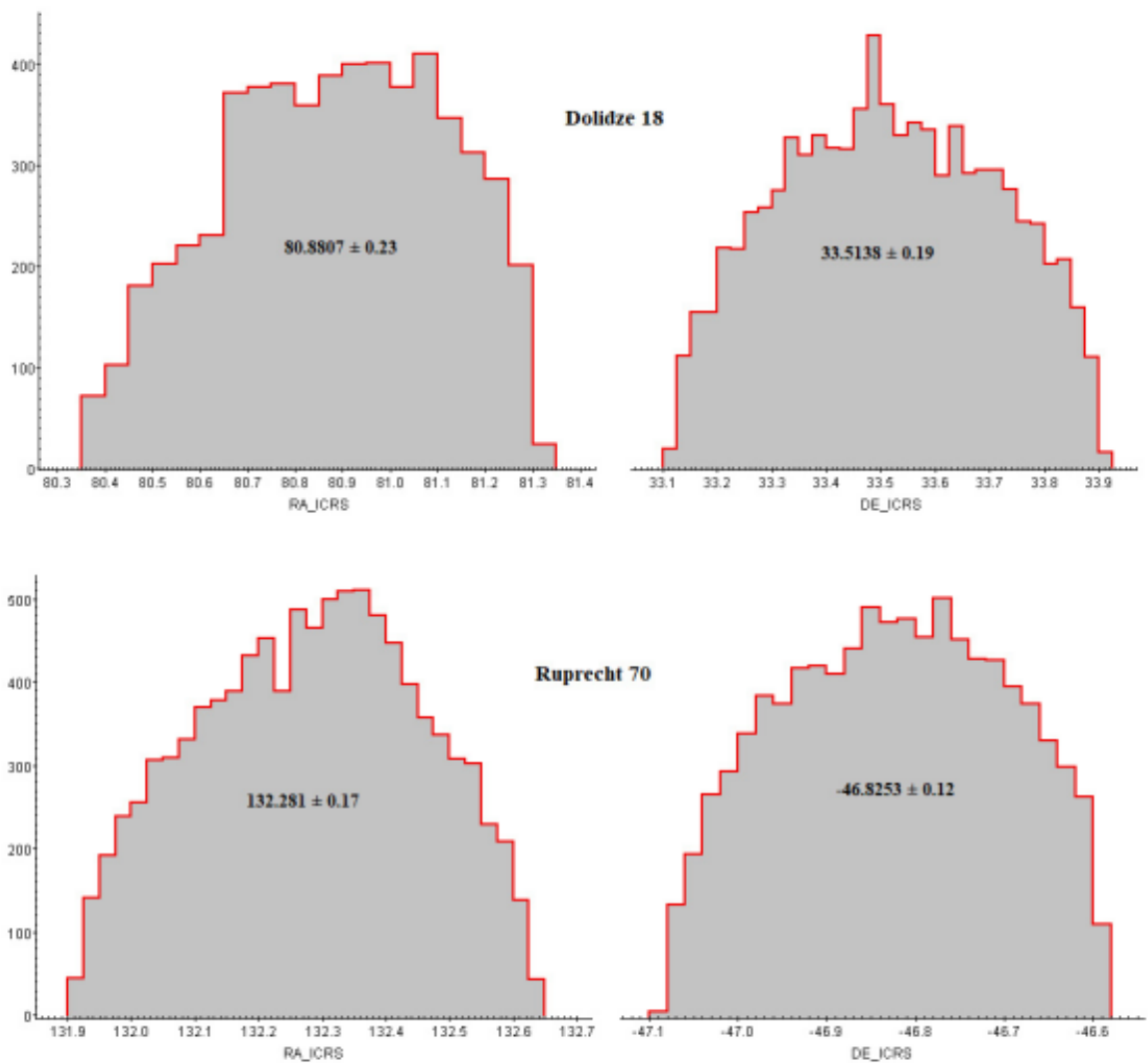

Fig. 5. Centers determination of the clusters Dolidze 18 and Ruprecht 70. The mean values of RA and DE are estimated in degrees for each cluster, which is found to be in good agreement with Dias et al. (2014), see Table 1.

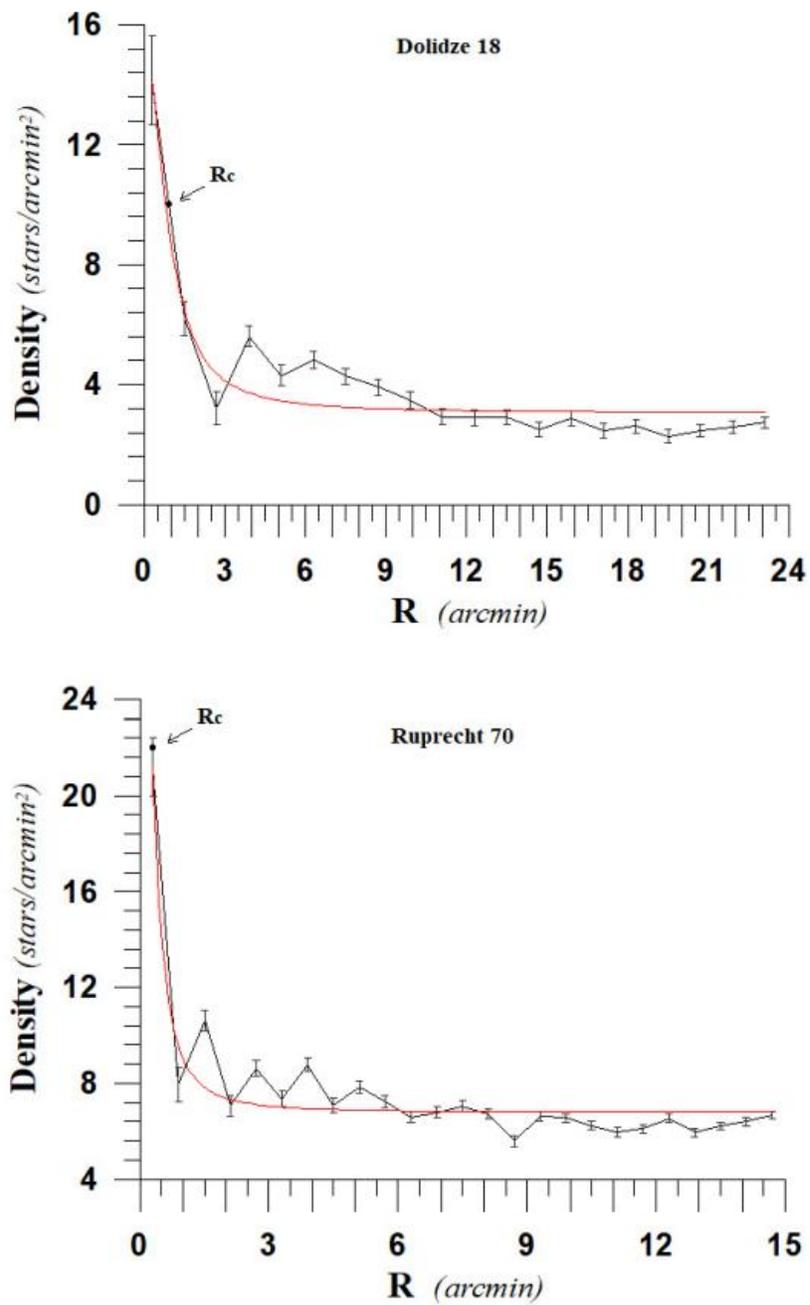

Fig. 6. Radial density profiles of the clusters Dolidze 18 and Ruprecht 70. The error bars refer to the Poisson distribution, and the red line refers to the fitted profile of King (1966). The limited radius is found to be 11.0 and 6.5 arcmin, respectively. $R_c$ refers to the core radius of the cluster. $R_c$ = 0.94 and 0.28 arcmin for Dolidze 18 and Ruprecht 70, respectively.

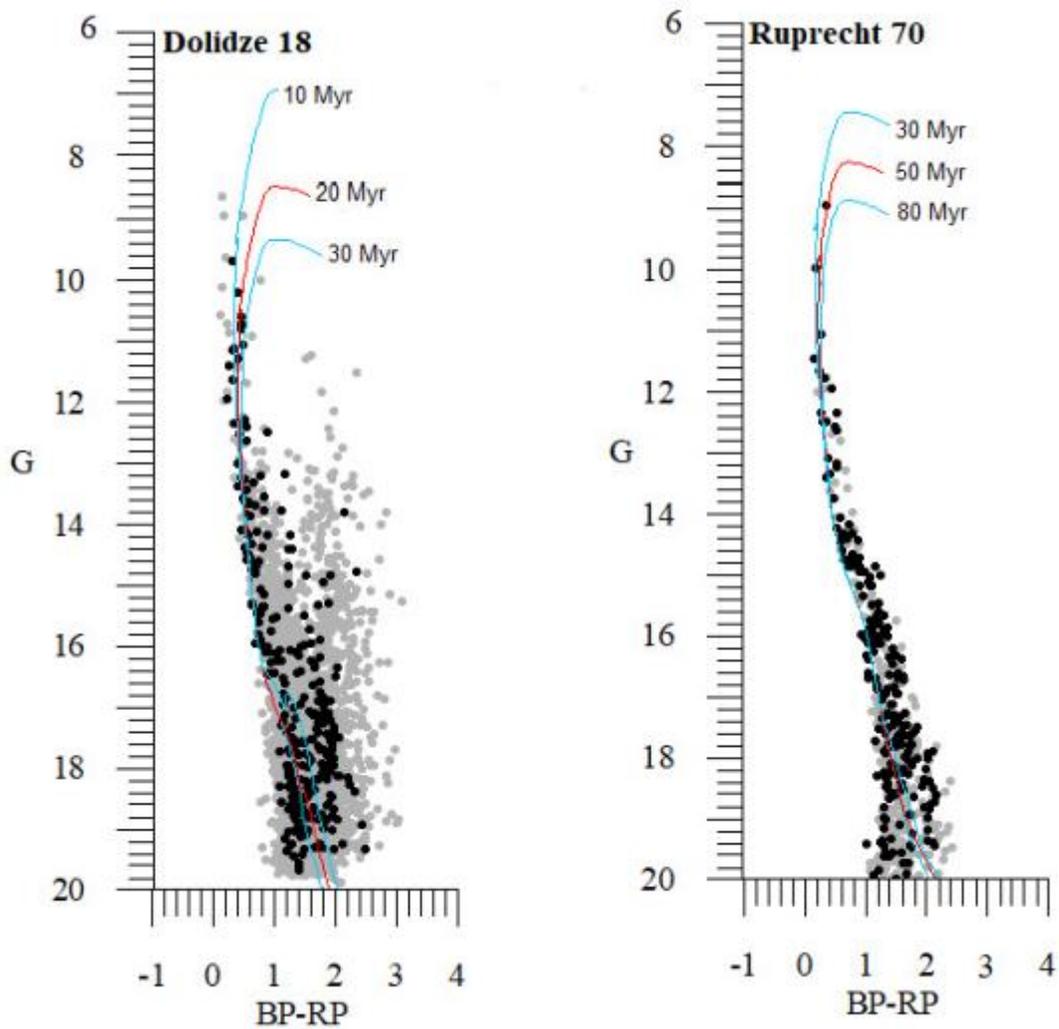

Fig. 7. The theoretical isochrones of the solar metallicity Z=0.0152 of Padova with three different ages have been applied to each cluster. The main parameters of Dolidze 18 and Ruprecht 70 are verified with ages= 20±10 and 50±10 Myr, distance moduli $m - M$=14.60±0.50 and 12.85±0.35 mag, and color excesses $E(BP - RP)$= 0.80±0.11 and 0.58±0.08 mag, respectively.

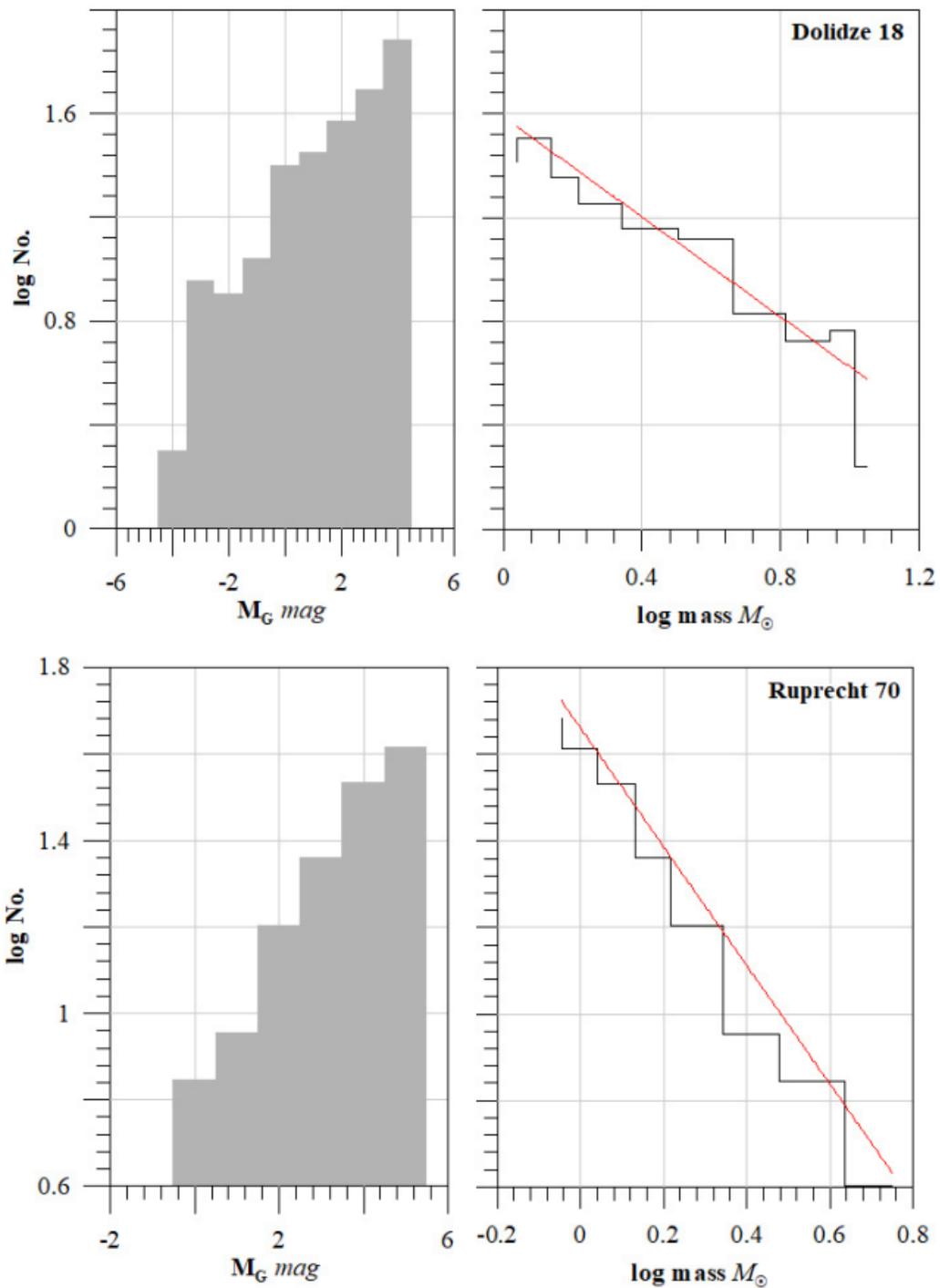

Fig. 8. The left-hand panels represent the luminosity function distribution, while the right-hand panels represent the mass function distribution of the two clusters. The red linear fit shows the slope of the relation, where $\alpha = -1.20$ and $-2.40$, respectively.

Table 1. The current estimate of the clusters' parameters.

| Parameter | Dolidze 18 | Ruprecht 70 |
|---|---|---|
| $\alpha(2000)$  $deg$ | 80.8807 | 132.281 |
| $\delta(2000)$  $deg$ | 33.5138 | -46.8253 |
| $l(2000)$  $deg$ | 173.5937 | 266.3256 |
| $b(2000)$  $deg$ | -1.4869 | -1.9140 |
| $Age$  $Myr$ | 20±10 | 50±10 |
| $E(BP-RP)$  $mag$ | 0.80±0.11 | 0.58±0.08 |
| $(m-M)$  $mag$ | 14.60±0.50 | 12.85±0.35 |
| $E(B-V)$  $mag$ | 0.62±0.11 | 0.45±0.08 |
| $A_G$  $mag$ | 1.65 | 1.20 |
| $A_V$  $mag$ | 1.90 | 1.40 |
| $(m-M)_o$  $mag$ | 12.95±0.60 | 11.65±0.45 |
| $Dist.$  $pc$ | 3890±180 | 2140±100 |
| $Plx.$  $mas$ | 0.28±0.12 | 0.45±0.10 |
| $pmRA$  $mas/yr$ | -0.21±0.06 | -5.19±0.15 |
| $pmDE$  $mas/yr$ | -1.41±0.13 | 4.63±0.12 |
| $Mem.$  $stars$ | 306 | 216 |
| $R_{lim}$  $arcmin$ | 11.0 | 6.5 |
| $R_c$  $arcmin$ | 0.94 | 0.28 |
| $C$ | 1.07 | 1.37 |
| $R_g$  $kpc$ | 12.21 | 8.74 |
| $f_{bg}$ | 3.0 | 6.8 |
| $f_o$ | 12.2 | 31.2 |
| $X_\odot$  $pc$ | 3863 | 137 |
| $Y_\odot$  $pc$ | 435 | -2135 |
| $Z_\odot$  $pc$ | -103 | -72 |
| $R_t$  $pc$ | 13.60 | 10.0 |
| $IMF slope$ | -1.20 | -2.40 |
| $Total$  $lumin.$  $mag$ | -8.7 | -5.8 |
| $Total$  $mass$  $M_\odot$ | 800 | 325 |
| $T_R$  $Myr$ | 55 | 14 |
| $\tau$ | 0.36 | 4.0 |